# Current-Driven Symmetry Breaking and Spin-Orbit Polarization in Chiral Wires

*Uiseok Jeong[1], Daniel Hill[2], Esmaeil Taghizadeh Sisakht[1], Binghai Yan[3], Angel Rubio[4], Carsten A. Ullrich[2], Noejung Park[1,4*]*

[1]Department of Physics, Ulsan National Institute of Science and Technology(UNIST), UNIST-gil 50, Ulju-gun, 44919, Ulsan, Republic of Korea.

[2]Department of Physics and Astronomy, University of Missouri, 701 S College Ave, Columbia, 65211, Missouri, USA.

[3]Department of Physics, Pennsylvania State University, 201 Old Main, University Park, 16802, Pennsylvania, USA.

[4]Max Planck Institute for the Structure and Dynamics of Matter, Luruper Ch 149, Hamburg, 22607, Hamburg, Germany.

KEYWORDS: Spin-orbit dynamics, Current-induced symmetry breaking, rt-TDDFT, Current-induced spin selectivity, Chiral wire

ABSTRACT

The spin dynamics of electrons in chiral molecular systems remain a topic of intense interest, particularly regarding whether geometric chirality inherently induces spin polarization in current-carrying electrons. In this work, we employ ab initio real-time time-dependent density functional theory (rt-TDDFT) to directly simulate the interplay between charge current, spin, and orbital. This real-time tracking extends beyond perturbative treatments, and we analyze how nonequilibrium currents effectively lift the symmetry constraints of screw rotation and time-reversal symmetry. We find that the emergence of spin and orbital angular momenta is dynamically correlated with a concomitant loss of translational (linear) momentum, which we interpret as an intrinsic consequence of current-driven symmetry lowering. The implications of this mechanism for chirality-induced spin selectivity and spintronic device design are discussed.

## INTRODUCTION

Charge current is not only the most essential operational component of electronic devices, but also one of the most common and experimentally accessible ways to drive condensed-matter systems out of equilibrium[1-3]. Under a biased current, the internal degrees of freedoms – such as spin and orbital angular momentum – undergo reorganization or transport[4, 5], leading to a rich variety of current-induced phenomena that have received growing interest in recent years[6-8]. For example, the electrically driven angular-momentum transport in spin Hall and orbital Hall systems has become a central topic of condensed-matter physics, forming the underlying constituents of what are now known as spintronics and orbitronics[3, 7, 9-16]. At reduced dimensionality, especially at surfaces and interfaces where inversion symmetry is broken, charge-to-angular-momentum conversion has been extensively analyzed within the linear perturbative regime through (Rashba-)Edelstein and orbital Edelstein effects[7, 17-19]. On the other hand, it has been proposed that the geometrical chirality of a one-dimensional transport channel would filter the spin state of charge carriers, a phenomenon known as chirality-induced spin selectivity (CISS)[20-24]. Considering these developments across diverse fields, a systematic investigation of current-carrying states and of how electric currents affect internal degrees of freedom are timely required.

In the present work, we particularly focus on a one-dimensional chiral wire, and through direct real-time, *ab initio* simulations, we show how an applied current, combined with its intrinsic geometric chirality, generates orbital and spin dynamics. To date, the interconversion between linear momentum and spin or orbital angular momentum has been primarily analyzed within linear-response theory. Likewise, the CISS effect has most often been investigated using perturbative approaches based on the nonequilibrium Green′s function (NEGF) formalism[21, 25-27]. In these fields, a full *ab initio* study of the effect of an applied external bias and the consequent current, beyond the linear regime, remains limited[28]. To address this gap, we employ real-time time-dependent density functional theory (rt-TDDFT) to simulate the response of chiral materials under applied electric fields, enabling real-time tracking of key physical observables such as charge current, orbital angular momentum, and spin angular momentum. We pay particular attention to how the intrinsic symmetry constraints of chiral chains are modified—or effectively lifted—in time-evolving nonequilibrium states driven by an increasing charge current.

We begin by examining how these constraints are encoded in the Hamiltonian and how their modification is reflected in the properties of static eigenstates of the chiral system. We then demonstrate that such constraints can be released dynamically in current-carrying states, even without any modification of the underlying Hamiltonian. Notably, Kramers' theorem enforces degeneracy of spinful Bloch states at time-reversal-invariant momenta for static eigenstates. During time evolution, however, this degeneracy is not necessarily preserved, even if the governing Hamiltonian remains time-reversal symmetric. In our real-time TDDFT calculations, we control the temporal profile of the external potential to disentangle the effect of the current from that of the driving field. Using *ab initio* real-time TDDFT simulations of a Se-based trigonal wire, we show that spin and orbital angular momenta are induced purely by the current, even after the external electric bias is switched off once charge flow has been initiated. This study establishes a framework for exploring current-induced polarization effects in chiral systems at the real-time quantum mechanical level, which could guide future investigations into electrically driven spintronic functionalities.

## RESULTS AND DISCUSSION

An ideal chiral wire can possess screw rotational symmetry, which involves a partial translation combined with a partial rotation, as schematically illustrated in Fig. 1. The corresponding unitary symmetry operator for the trigonal chiral wire can be given by:

$$\hat{Q} = \hat{T}_{a/3}\hat{R}_{2\pi/3} = \hat{R}_{2\pi/3}\hat{T}_{a/3} \quad \text{Eq. 1}$$

When the eigenvalue spectrum is non-degenerate, the screw rotation unitary symmetry operator ($\hat{Q}$) in Eq. 1 shares the same eigenstates as the Hamiltonian itself. Consequently, in this example of the trigonal helical wire, the in-plane spin expectation value of each Bloch eigenstate must be invariant under a $C_3$ rotation, which requires the in-plane spin components to vanish.

$$\begin{gathered} \hat{Q}\left|\psi\right\rangle = \lambda\left|\psi\right\rangle,\ \left|\lambda\right|^2 = 1, \\ \left\langle \psi_{n,k} \mid \hat{Q}^{\dagger}\hat{\mathbf{S}}\hat{Q} \mid \psi_{n,k} \right\rangle = \left\langle \psi_{n,k} \mid \hat{R}^{\dagger}_{2\pi/3}\hat{\mathbf{S}}\hat{R}_{2\pi/3} \mid \psi_{n,k} \right\rangle, \quad \text{Eq. 2} \\ \left\langle \psi_{n,k} \mid \hat{S}_x \mid \psi_{n,k} \right\rangle = 0,\ \left\langle \psi_{n,k} \mid \hat{S}_y \mid \psi_{n,k} \right\rangle = 0. \end{gathered}$$

Note that the first line of Eq. 2 indicates that the in-plane spin components ($S_x, S_y$) must be equal to their values after $2\pi/3$ rotation, which is only allowable if they are identically zero, as explicitly stated in the second line. As a real-material example of the screw-symmetric system, we performed the first-principles static density functional theory (DFT) calculations of a trigonal selenium (Se) wire[29-31]. Figure 2 summarizes the results for the Se wire. Details of the computational methods are provided in the Method section. The Se wire exhibits substantial spin-orbit coupling, resulting in energy band structures that reflect a distinctive spin-momentum locking structure, as presented in Fig. 2(a). For each Bloch state ($\psi_{n,k}$), we calculated the expectation values of spin components and the *z*-component of orbital angular momentum ($\hat{L}_z = xp_y - yp_x$). For this one-dimensional structure, the in-plane coordinates $(x, y)$ are well-defined, and the orbital angular momentum was evaluated explicitly without any approximation scheme. Note that $L_x$ and $L_y$ are not well defined because the *z*-coordinate is not definitely determined in the 1-dimensional structure. In accordance with the symmetry argument presented above in Eq. 1, the in-plane spin components of each Bloch eigenstate vanish, while the *z*-components of spin and orbital degrees of freedom are prominently manifested depending on the Bloch momentum, as presented in Fig. 2(a).

Next, we explore symmetry breaking by applying a static electric field perpendicular to or along the wire axis, as illustrated in the top panel of Fig. 2(b) and (c), respectively. This applied field breaks the rotational symmetry, and consequently, the screw symmetry no longer serves as a governing constraint. As a result, each Bloch eigenstate now exhibits non-vanishing in-plane spin expectation values ($S_{x,y} \neq 0$), as indicated by the color scheme in the second panel of Fig. 2(b). The screw symmetry Eq. 1 can also be removed by breaking the translation symmetry, as depicted in the top panel of Fig. 2(c). In this case, the axial electric field additionally incorporates the scalar potential $-E_z(t)z$, which explicitly breaks translational symmetry, rendering the Hamiltonian incompatible with the screw rotation: $\hat{H}\hat{Q} \neq \hat{Q}\hat{H}$. Explicit formal proof of screw-symmetry breaking by the axial external electric field is summarized in Supplementary Note 1. In the velocity gauge, we incorporate the uniform constant electric field along the $z$-direction as a time-dependent vector potential, and the Hamiltonian is now time-dependent, as summarized in Eq. 3[32-35]. The time-evolution scheme and the functional details of the rt-TDDFT are provided in the Method section.

$$
\begin{gathered}
i\hbar\frac{\partial}{\partial t}\psi_{n,k}(\mathbf{r},t)=\hat{H}(\mathbf{r},t)\psi_{n,k}(\mathbf{r},t)\\
\hat{H}(\mathbf{r},t)=\frac{1}{2m}\left(-i\hbar\nabla+\frac{e}{c}\mathbf{A}_{ext}(t)\right)^2+\sum_{\lambda}V_{atom}(\mathbf{r}-\mathbf{R}_{\lambda})+V_{DFT}\left[\rho(\mathbf{r},t)\right]+V_{SOC}\\
\mathbf{E}_{ext}=E\hat{z}=-\frac{1}{c}\frac{\partial}{\partial t}\mathbf{A}_{ext}
\end{gathered}
\quad \text{Eq. 3}
$$

We started with the static eigenstates and evaluated the transient instantaneous band-energy expectation values from the time-evolving state($\psi_{n,k}(t)$). The corresponding instantaneous spin and orbital expectation values are also computed, as given in Eq. 4. The time-translating band energies are plotted with respect to the initial label of the Bloch momentum ($k$), as depicted in the second and third panels of Fig. 2(c). In this example, the band structures almost rigidly shifted equivalently to a translation over the Brillouin zone by the applied vector field $k(t)=k+eA(t)/\hbar c$.

$$
\begin{gathered}
\varepsilon_{n,k}(t)=\left\langle\psi_{n,k}(t)\middle|\hat{H}(t)\middle|\psi_{n,k}(t)\right\rangle\\
\theta_{n,k}(t)=\left\langle\psi_{n,k}(t)\middle|\hat{\theta}\middle|\psi_{n,k}(t)\right\rangle,\ \hat{\theta}=\hat{S}_x,\hat{S}_y,\hat{S}_z \text{ or } \hat{L}_z
\end{gathered}
\quad \text{Eq. 4}
$$

The instantaneous band energy expectation($\varepsilon_{n,k}(t)$), as defined in Eq. 4, is plotted in Fig. 2(c), and the spin and orbital expectation values of each time-evolving Bloch state are depicted with color codes. Those time-translating instantaneous band energies, spin, and orbitals are plotted with respect to the initial label of the Bloch momentum($k$), in the second and third panels of Fig. 2(c).

We now examine the time profiles of the total current, spin, and orbital angular momentum of the system. To simulate a smooth metallic response, we introduced +0.1 hole doping into the unit cell of the 3 Se atoms. For electric bias, we chose five different strengths of constant electric field, ranging from 0.8mV/Å to 1.2mV/Å in increments of 0.1 mV/Å. As done above, all these electric fields were gradually turned on with a 30 fs ramping period to ensure numerical stability. The time profile of the applied electric field is shown in Fig. 3(a). As time goes on, physical observables were evaluated as functions of time from the expectation values of the time-evolved Bloch wavefunctions ($\psi_{n,k}(t)$), as defined in the second line of Eq. 4. For the total values of those observables, we summed them over the bands and Brillouin zone weighted by their occupation numbers, as follows:

$$
\begin{aligned}
I_z(t) &= -\frac{e}{m_e}\sum_{n,k} f_{n,k}\left\langle \psi_{n,k}(t)\middle|\hat{\pi}_z\middle|\psi_{n,k}(t)\right\rangle \\
L_z(t) &= \sum_{n,k} f_{n,k}\left\langle \psi_{n,k}(t)\middle|\hat{L}_z\middle|\psi_{n,k}(t)\right\rangle \qquad \text{Eq. 5} \\
S_z(t) &= \sum_{n,k} f_{n,k}\left\langle \psi_{n,k}(t)\middle|\widehat{S}_z\middle|\psi_{n,k}(t)\right\rangle
\end{aligned}
$$

Here, the $f_{n,k}$ denotes the occupation of the ground Bloch states, $e$ and $m_e$ indicates the charge and mass of the electron, respectively. In what follows, we use the gauge-invariant form of the orbital angular momentum defined by the cross product of position and the gauge-invariant momenta($\mathbf{r}\times\hat{\boldsymbol{\pi}}(t)$). Figures 3(b, c, d) present the time evolution of current, orbital angular momentum, and spin angular momentum along the chain axis. Before the time evolution initiated by the external electric bias, the system resides in the ground state that respects time-reversal symmetry, with the spin orbitals of the Bloch states forming Kramers' pairs. At this stage, contributions from states at opposite Bloch momenta cancel, resulting in a vanishing expectation value of spin polarization and preserving time-reversal symmetry. However, the unitary time-evolution operator does not commute with the antiunitary time-reversal operator. As a result, the Kramers' pairing between time-reversal partners is not necessarily preserved during the evolution, irrespective of whether the instantaneous Hamiltonian manifestly contains the time-reversal breaking term or not. Figures 3(c) and (d) explicitly demonstrate that the resulting current flow introduces effective time-reversal symmetry breaking. The transport of the initial spin-paired up and down states under the applied electric bias results in spin polarization along the wire axis (*z*-direction). These findings point toward a broader principle: in systems with geometric chirality and spin–orbit coupling, the application of charge current can induce time-reversal symmetry breaking, giving rise to spin and orbital angular momentum.

A detailed technical discussion is necessary for the time profiles of the in-plane components of spins ($S_x(t)$ and $S_y(t)$). As is always the case in solid-state problems with periodic boundary conditions, the choice of unit-cell coordinates is not unique and can be chosen arbitrarily. Recall that, at $t = 0$, the system was under the screw rotation symmetry with all vanishing in-plane spins, as described in Eq. 2. Over time, as the electric bias lowers the symmetry, the in-plane spins emerge, but noticeably, the time profiles of those in-plane spins depend on the arbitrarily selected unit-cell coordinate. This implies that the obtained values of these in-plane spins are not physically measurable. This can be attributable to the fact that the time-dependent Hamiltonian or the time-evolution unitary operator is commutative with the lattice translation. For example, one may repeat the calculation from the initial states with an arbitrarily shifted coordinate, for example, by $a/N$.

$$\begin{aligned} |\Psi(t)\rangle &\rightarrow \mathcal{T}\exp\left(\int_0^t \hat{H}(\tau)d\tau\right)\hat{T}_{a/N}|\Psi_{t=0}\rangle \\ &= \hat{T}_{a/N}\mathcal{T}\exp\left(\int_0^t \hat{H}(\tau)d\tau\right)|\Psi_{t=0}\rangle = \hat{T}_{a/N}|\Psi(t)\rangle. \end{aligned} \quad \text{Eq. 6}$$

Equation 6 tells that the time-evolved state, started from the shifted coordinates by $a/N$, can be obtained by translating the time-evolved state of the original coordinate. It is noteworthy that, for an arbitrary translation of this screw system, there must be a corresponding rotation operator, exactly equivalent to the translation, and consequently, the in-plane spins in the new coordinates are rotated by that amount.

$$\begin{aligned} &|\Psi(t)\rangle \rightarrow |\Psi'(t)\rangle = \hat{R}_{2\pi/N}|\Psi(t)\rangle, \\ &\langle\Psi'|\hat{\mathbf{S}}|\Psi'\rangle = \langle\Psi|\hat{R}^{\dagger}_{2\pi/N}\hat{\mathbf{S}}\hat{R}_{2\pi/N}|\Psi\rangle = \langle\Psi|\hat{\mathbf{S}}_{\text{rotated}}|\Psi\rangle. \end{aligned} \quad \text{Eq. 7}$$

Note that the second line of Eq. 7 tells that the time profile of the spins calculated with the primed coordinates is equal to that of the real-time spin obtained in the unprimed coordinates by $\hat{R}_{2\pi/3}$. As an example, we performed the rt-TDDFT calculations and compared the results with those shifted by $a/3$ and $2a/3$, as summarized in Figs. 4(a, b, c). At every moment, the in-plane spins ($S_x(t)$, $S_y(t)$) are rotated in exact agreement with Eq. 7, and thus the magnitude of the in-plane spin $\sqrt{S_x^2 + S_y^2}$ remains consistent across these arbitrary choices of unit cells. The axial component of the spin ($S_z(t)$) is not affected by these choices of unit cells, and our first-principles rt-TDDFT calculation indeed confirms the consistency of $S_z(t)$, irrespective of the coordinate choices. (Compare $S_z(t)$ in Figs. 4(a, b, c)).

As shown in Fig. 3, applying a uniform constant electric field along the axial direction of the Se trigonal wire generates a current that simultaneously triggers a sharp rise in spin and orbital angular momenta ($L_z, S_z$). In short, the angular momenta, frozen by the Kramers' pairing under time-reversal symmetry of the initial static Hamiltonian, are now released upon the application of the axial electric bias. In realistic devices, the current flow is primarily driven by chemical potential differences between the leads, with the electric field being largely confined near the contacts. As a result, even when a finite or nearly steady current dominates in the channel region, the external field is mostly screened out, and the main portion of the transport channel may remain effectively free of the bias field. This raises the question of whether the time-reversal symmetry-breaking features observed in Fig. 3(c, d) can persist—or even emerge—solely due to the presence of current, in the absence of an external bias field. To examine this question, we realize the constant current states without external bias in our scheme of real-time TDDFT.

The external electric field $E_{ext}$ was gradually turned on and reached a constant value after a ramping period of 30 fs. After a finite duration, the external field was gradually terminated, as depicted in Fig. 5(a). Two different durations of the electric bias ($E_1(t)$ and $E_2(t)$) are presented in Fig. 5, terminating at 120 fs and 180 fs, respectively. We also tested various other field strengths and durations; two representative cases are presented in Fig. 5, with the remaining results summarized in the Supplementary Information. When the applied uniform electric field is turned off, the external vector potential, as defined in the third line of Eq. 3, becomes constant.

Usually, a constant in the potential can, in principle, be gauged out. However, in the case of a gradually ramped electric field, the gauge is fixed at the initial time ( $t = 0$ ), and the constant vector potential that remains after the field is turned off lies within this initially chosen gauge. Therefore, the constant potential is not arbitrary but has physical significance.

Due to the work done by the external bias field, the total energy of the system increases monotonically while the field is applied. After the external electric field is turned off, the total energy remains constant, as shown in Fig. 5(b). A formal proof of this energy conservation, after switching off the external electric field, is provided and discussed in detail in the Supplementary Note 2. As above, we calculated the instantaneous expectation values of the band energy, $\varepsilon_{n,k}(t) = \left\langle \psi_{n,k}(t) \middle| \hat{H} \middle| \psi_{n,k}(t) \right\rangle$, Fig. 5(c) shows that the band energies exhibit negligible evolution over time once the external bias is switched off. Compare Fig. 2(c), the band energies under the external field, and Fig. 5(c) after the electric bias was turned off. Our rt-TDDFT simulation does not include any explicit dissipation mechanism. As a result, even after the external electric field is turned off, the dynamics of the charge current, spin, and orbital angular momentum persist, while the total energy is fixed to the value of the last moment of the external electric field. These ongoing dynamics represent the internal dynamics of the system (charge current, spin, orbital) under the constraint of a conserved total energy. Fig. 5(d, e) shows that those rotational degrees of freedom—particularly the spin—respond more slowly. When the electric bias is terminated at 120 fs, the spin and orbital angular momenta remain nearly inactive. When the electric field was turned off at t=120fs, not only does the total energy but also the current remain constant, as presented in Fig. 5(d). However, when the external bias persists for a longer duration, a greater amount of current is induced, which effectively triggers spin and orbital responses. For example,

Fig. 5(e) illustrates the sharp emergence of spin and orbital dynamics in the case of the field profile $E_2(t)$.

It is noteworthy that, as the rotational degrees of freedom become activated, the current begins to decrease while the total energy remains conserved. Viewed conversely, this observation suggests that spin and orbital angular momenta emerge at the expense of linear momentum. In other words, the sharp increase in spin and orbital components observed in Fig. 5(e) can be attributed to the conversion of linear momentum into angular momentum via spin–orbit coupling. We conjecture that the conversion from linear to angular momentum could generally be observed when spin-orbit-coupled chiral wires are subjected to a biased current. These findings indicate that the presence of a current indeed serves as a source for activating spin and orbital degrees of freedom. As presented in Fig. 5(e), when the current does not reach a certain threshold value $I_{th}$, the spin and orbital components remain frozen, and the Kramers' pairing remains seemingly intact. As detailed in Fig. S5 in the Supplementary Information, we systematically varied both the field duration of the applied bias and found that it is the total amount of induced current—rather than the field strength or duration alone—that governs the onset of spin and orbital responses. This current-dependent behavior suggests the existence of a threshold, beyond which linear momentum is effectively converted into angular momentum through spin–orbit interaction.

## CONCLUSIONS

Using *ab initio* real-time time-dependent density functional theory calculations, we investigated how an externally applied bias field and charge current drive spin and orbital angular momentum dynamics in a screw-symmetric one-dimensional chiral wire. We specifically focused on how a net spin and orbital angular momentum polarization emerge beyond the constraint of screw-rotation symmetry and time-reversal Kramers' degeneracy. Through the real-time tracking of the evolutions of charge current, spin, and orbital, we demonstrated that net orbital angular momenta arise at the expense of linear momenta. We suggest that the emergence of net spin in dynamical states is effectively attributable to current-driven time-reversal symmetry breaking, rather than an alteration of symmetry at the level of the Hamiltonian. Overall, our first-principles study highlights that the net spin momenta observed in the current-flowing states—such as those found in CISS devices— arise from the intrinsic quantum mechanical interplay between charge current, spin, and orbital angular momenta on a SOC-equipped chiral geometry.

## Methods

### Static DFT calculations for the ground state electronic structures

The static ground-state electronic structure was obtained using standard density functional theory (DFT) calculations. We employed the Quantum ESPRESSO package[36, 37] to achieve self-consistently converged electronic states. The exchange-correlation energy was treated within the Perdew-Burke-Ernzerhof (PBE)[38] form of the generalized gradient approximation (GGA). Fully relativistic, norm-conserving pseudopotentials were used to account for the atomic potential and spin-orbit coupling effects. The energy cutoff for the plane-wave basis set is set to 60 Ry. To model an isolated single wire, a sufficiently large vacuum of 10Å was used in the in-plane (*x, y*) directions. To apply electric fields in the *x* and *y* directions as Fig. 1(b), we employed the homogeneous finite-field approach[39, 40]. The angular momentum of each eigenstate was explicitly calculated using Eq. 4 from the time-independent wavefunctions.

### Real-time time-dependent density functional theory calculations

To simulate the biased situations and the charge current therein, a uniform static electric field was applied along the wire axis. In the velocity gauge, the electric field is represented by a time-dependent vector potential, as given below:

$$
\begin{gathered}
i\hbar\frac{\partial}{\partial t}\psi_{n,\mathbf{k}}(\mathbf{r},t)=\hat{H}(\mathbf{r},t)\psi_{n,\mathbf{k}}(\mathbf{r},t)\\
\hat{H}(\mathbf{r},t)=\frac{1}{2m}\hat{\boldsymbol{\pi}}^2+\sum_{\lambda}V_{atom}(\mathbf{r}-\mathbf{R}_{\lambda})+V_{DFT}\left[\rho(\mathbf{r},t)\right]+\frac{\hbar}{4m^2c^2}\boldsymbol{\sigma}\cdot\nabla U\times\hat{\boldsymbol{\pi}} \quad \text{Eq. 8}\\
\hat{\boldsymbol{\pi}}=-i\hbar\nabla+\frac{e}{c}\mathbf{A}_{ext}(t)
\end{gathered}
$$

In this equation, ($n$, $\mathbf{k}$) denotes the band index and the Bloch wave vector, while $V_{atom}(\mathbf{r}-\mathbf{R}_{\lambda})$ represents the pseudopotential of the $\lambda$-th atom. In the second line in Eq. 8, $V_{DFT}\left[\rho(\mathbf{r},t)\right]$ collectively includes the density-dependent potentials, such as the Hartree and exchange-correlation contributions. Throughout this work, the adiabatic approximation is employed, and no explicit treatment of the memory effect is considered. For the time integration, we employed the Crank–Nicolson time evolution method for its numerical stability and computational efficiency. In these simulations, the discretized time step ($\Delta t$) was set to 2.414 attoseconds.

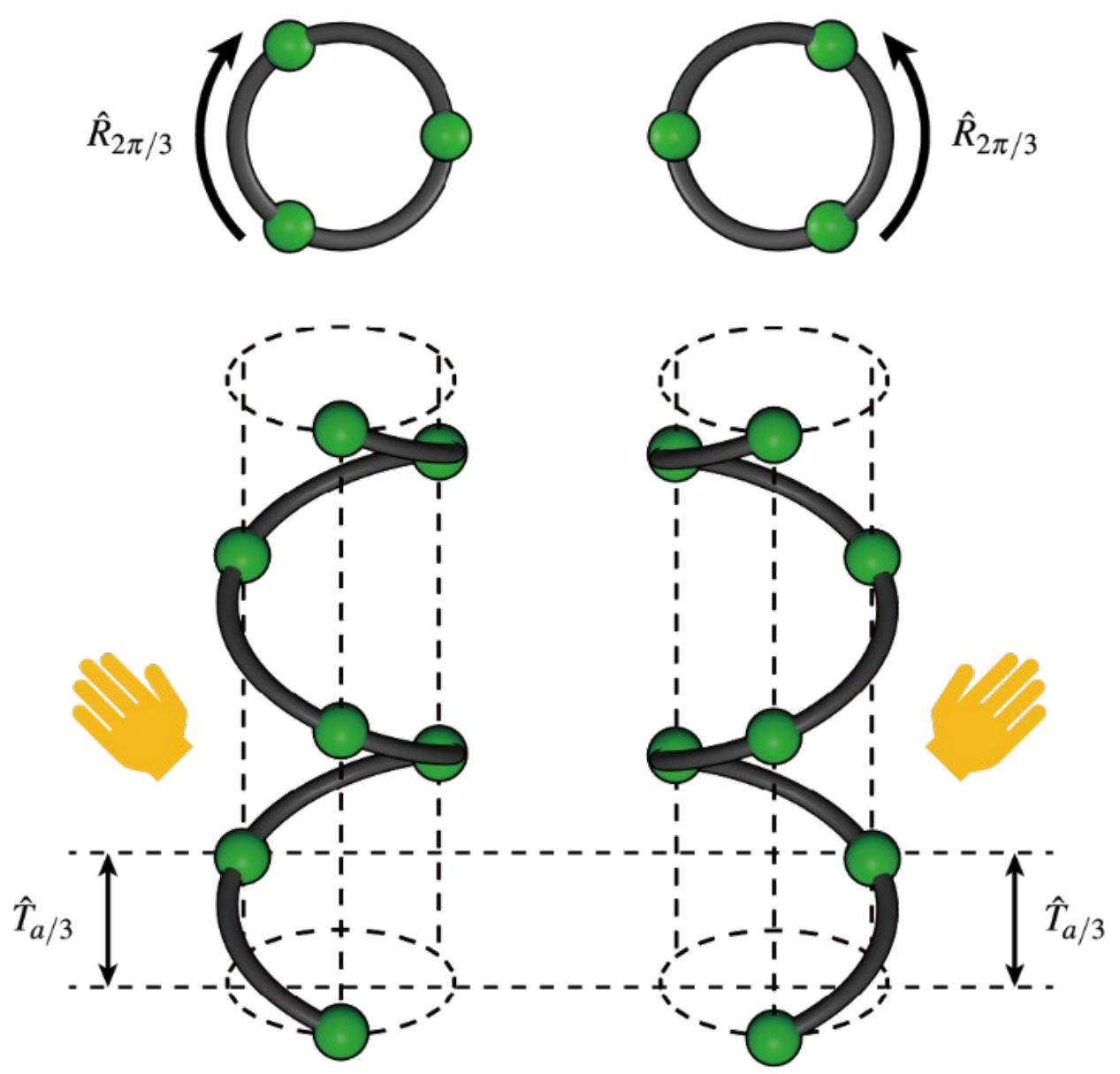

**Figure 1**. Schematic illustration of screw rotation symmetry. The top and bottom panels show the top and side views, respectively. The left and right columns represent left-handed and right-handed chains, denoted by the corresponding hand symbols.

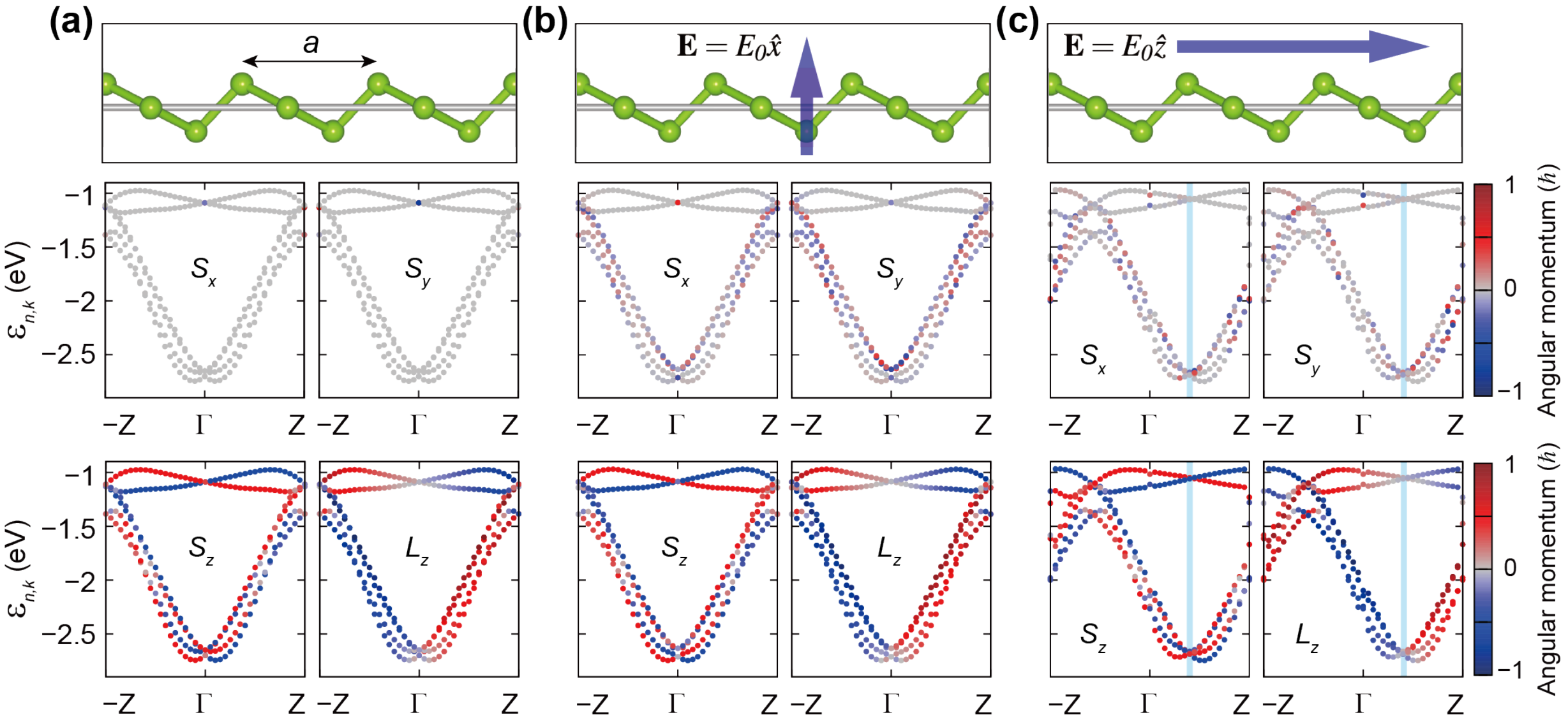

**Figure 2**. The electronic structure and spin-orbital texture of the Se trigonal one-dimensional wire (a) under the constraint of the screw-rotation symmetry; and with the screw symmetry being released by the (b) perpendicular and (c) axial bias field, respectively. (a) and (b) are static DFT calculations, whereas for (c), the time-evolutions of the Bloch states are obtained in the velocity gauge: the shown band energy expectations are taken at a particular time snapshot. The vertical sky blue lines in (c) are a guide to the eye for the instantaneously shifted Bloch momentum (see the main text).

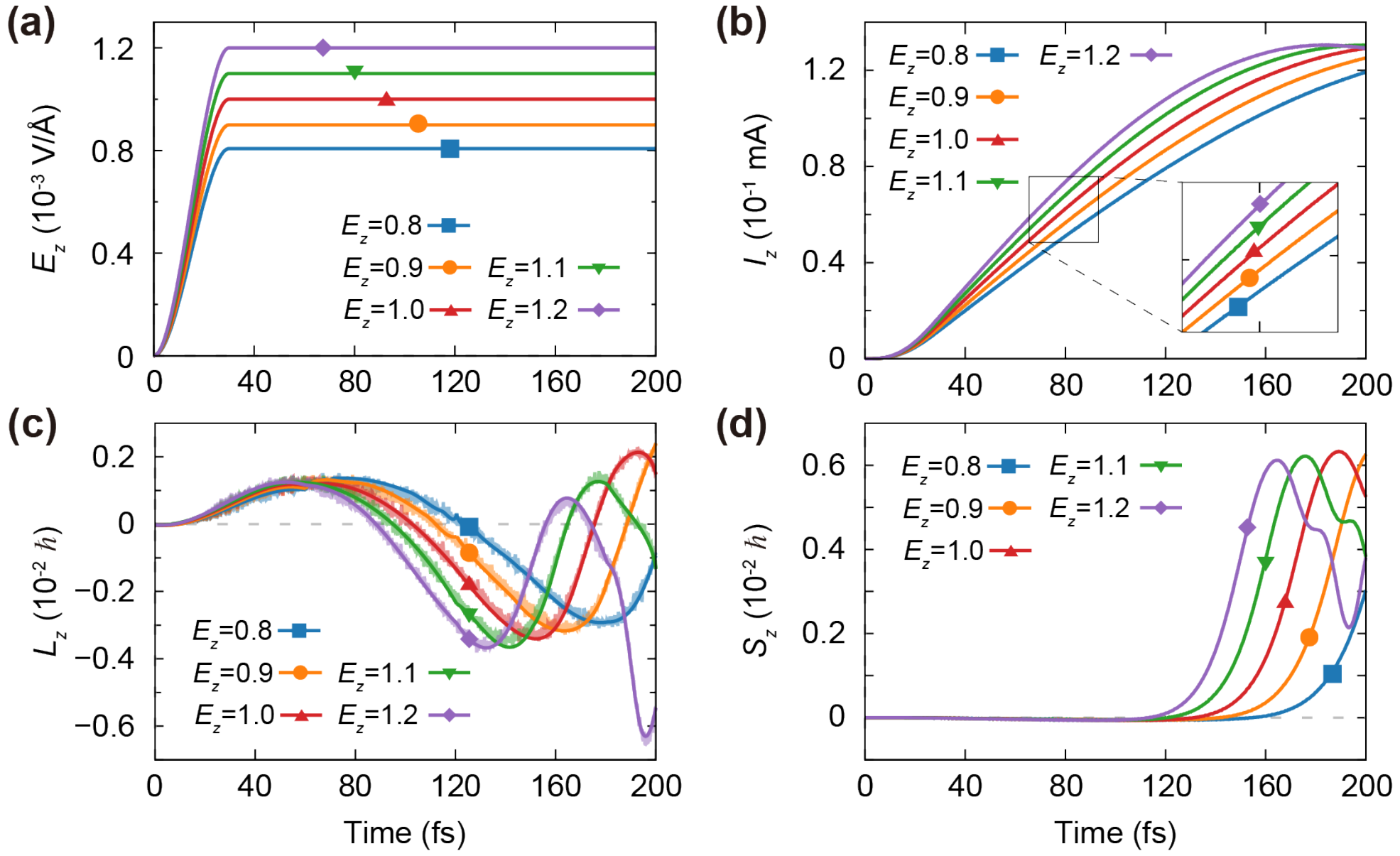

**Figure 3**. Time evolution of observables from rt-TDDFT simulations. Time-dependent profiles of (a) the applied electric bias, (b) the obtained charge current, (c) the orbital angular momentum, and (d) the spin angular momentum along the chain axis, respectively. Colored markers indicate different electric field strengths in unit of mV/Å

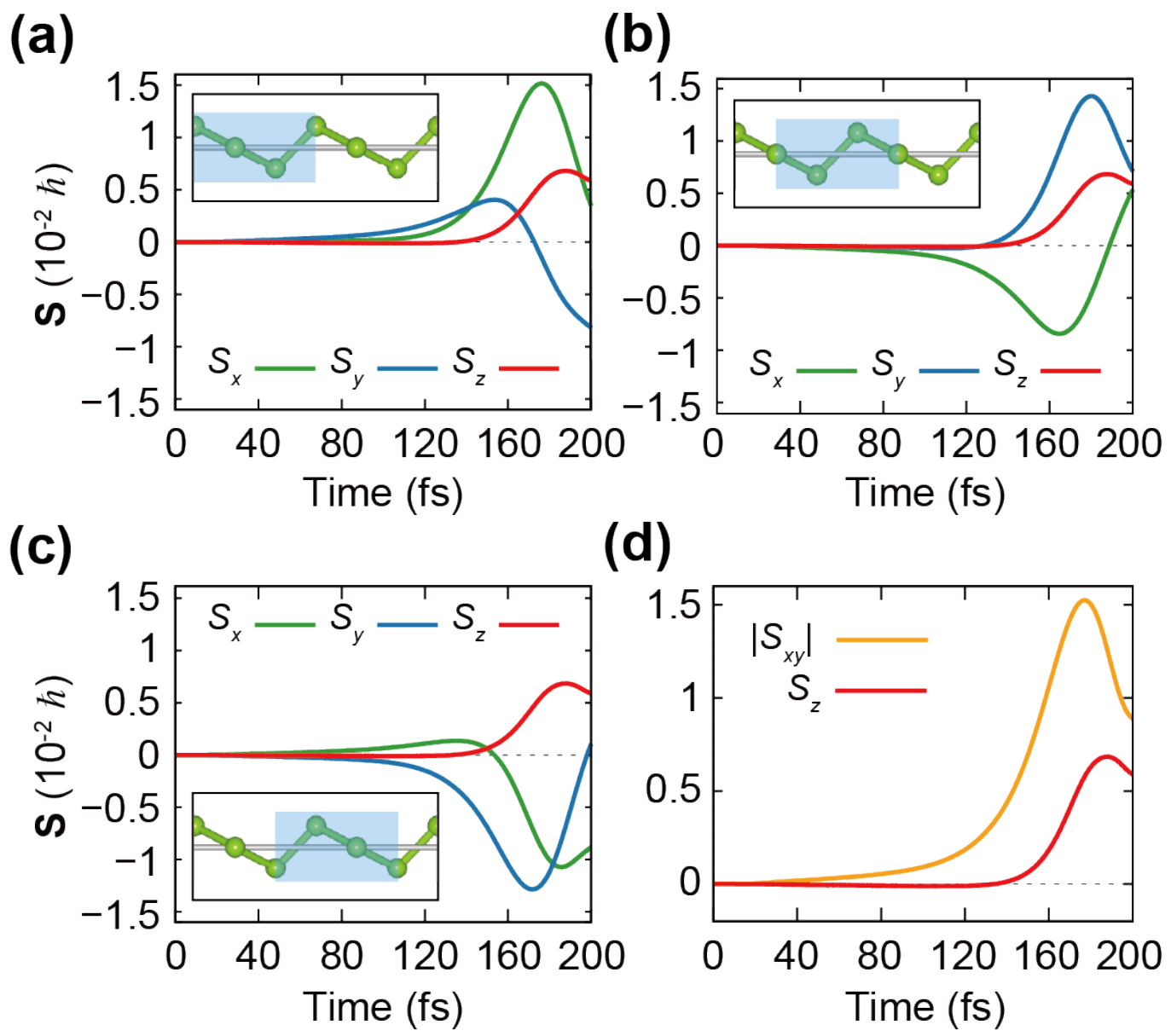

**Figure 4**. Arbitrariness of the in-plane spin depending on the choice of unit-cell coordinates. The unit-cell for (b) and (c) is shifted by $a/3$ and $2a/3$, respectively, compared with that of (a), along the wire axis. Time profiles of $S_x(t)$ and $S_y(t)$ are presented, and insets depict the unit-cell selections. Across all these calculations, the z-component of the spin and the magnitude of the in-plane spin ( $S_z(t) = (S_x^2 + S_y^2)^{1/2}$ ) are invariant over the arbitrary choice of the unit-cell, as summarized in (d).

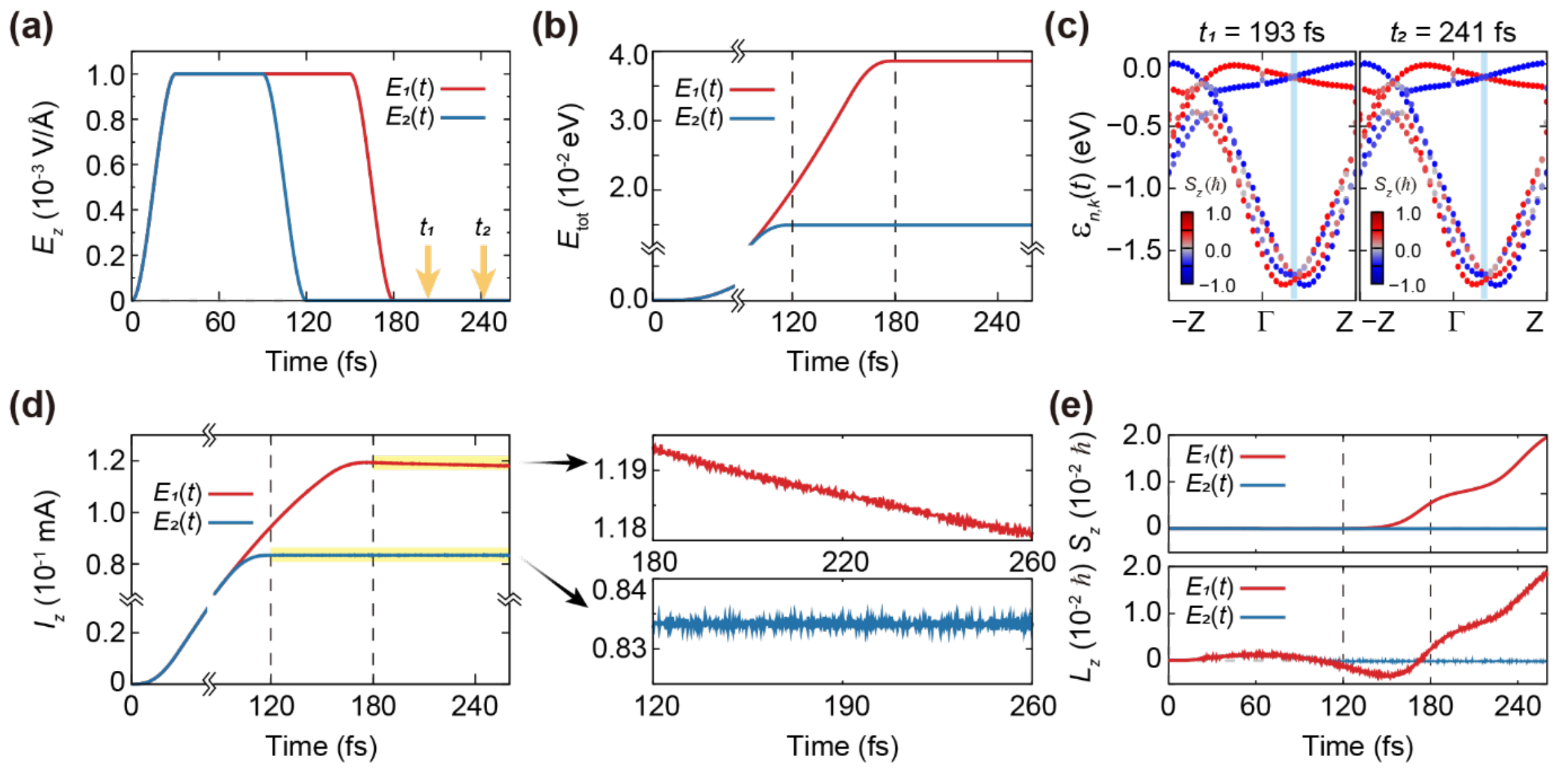

**Figure 5**. Time evolution of band energy, charge current, spin, and orbital angular momentum in the Se trigonal chiral wire under a finite-duration electric field. (a) Time profiles of the applied electric fields with two different durations: $E_1(t)$ and $E_2(t)$ terminate at 120 fs and 180 fs, respectively. (b) The time profiles of the total energies. (c) Instantaneous band energy expectations, $\varepsilon_{n,k}(t)$, are evaluated after the electric bias is turned off. Time profiles of (d) charge current, (e) spin($S_z$), and orbital angular momentum($L_z$) along the chain axis. Vertical dashed lines in (b, d, e) indicate the time of 120 and 180 fs from which $E_1(t)$ and $E_2(t)$ are turned off, respectively. The total energy in (b) is referenced to the initial value at $E_{tot}(t = 0) = -76.63$eV per unit cell. The snapshot times for the band energies shown in (c) ($t_1$ and $t_2$) are indicated with downward arrows in (a).

## ASSOCIATED CONTENT

Jeong, U.; Hill, D.; Sisakht, E. T.; Yan, B.; Rubio, A.; Ullrich, C. A.; Park, N. Current-Driven Symmetry Breaking and Spin-Orbit Polarization in Chiral Wires. **2025**, 2508.19519. *arXiv*. https://arxiv.org/abs/2508.19519 (accessed August 27, 2025).

**Supporting Information.** The following file is available free of charge at;

Note S1: Screw-rotation symmetry operator and symmetry breaking by an axial electric field; Note S2: Conservation of total energy after field termination; Note S3: Time evolution under a fixed ground-state density-functional Hamiltonian; Note S4: Effects of chirality and spin–orbit coupling on momentum conversion; Figure S1: Influence of self-consistency in the exchange-correlation potential; Figure S2: Comparison of linear-to-angular momentum conversion in chiral and nonchiral chains; Figure S3: Electronic structure under a transverse field; Figure S4: Unit-cell dependence of spin angular momentum; Figure S5: Additional calculations for varying field durations; Figure S6: Magnified view of the current in Figure S5.

## AUTHOR INFORMATION

**Corresponding Author**

*noejung@unist.ac.kr

https://orcid.org/0000-0002-2359-0635

Author Contributions

The manuscript was written through the contributions of all authors. All authors have given approval to the final version of the manuscript.

## ACKNOWLEDGMENT

U.J., E.T.S., and N.P. were supported by the National Research Foundation of Korea (NRF) grants funded by the Korean government (MSIT) (No. RS-2023-00257666, RS-2023-00208825), and by a KIAT grant funded by the Korean government (MOTIE) (P0023703, HRD Program for Industrial Innovation). C.A.U. was supported by the U.S. Department of Energy, Office of Science, Office of Basic Energy Sciences, under Award No. DE-SC0019109. B.Y. acknowledges the financial support by the Penn State Materials Research Science and Engineering Center for Nanoscale Science (MRSEC) under National Science Foundation award DMR-2011839.

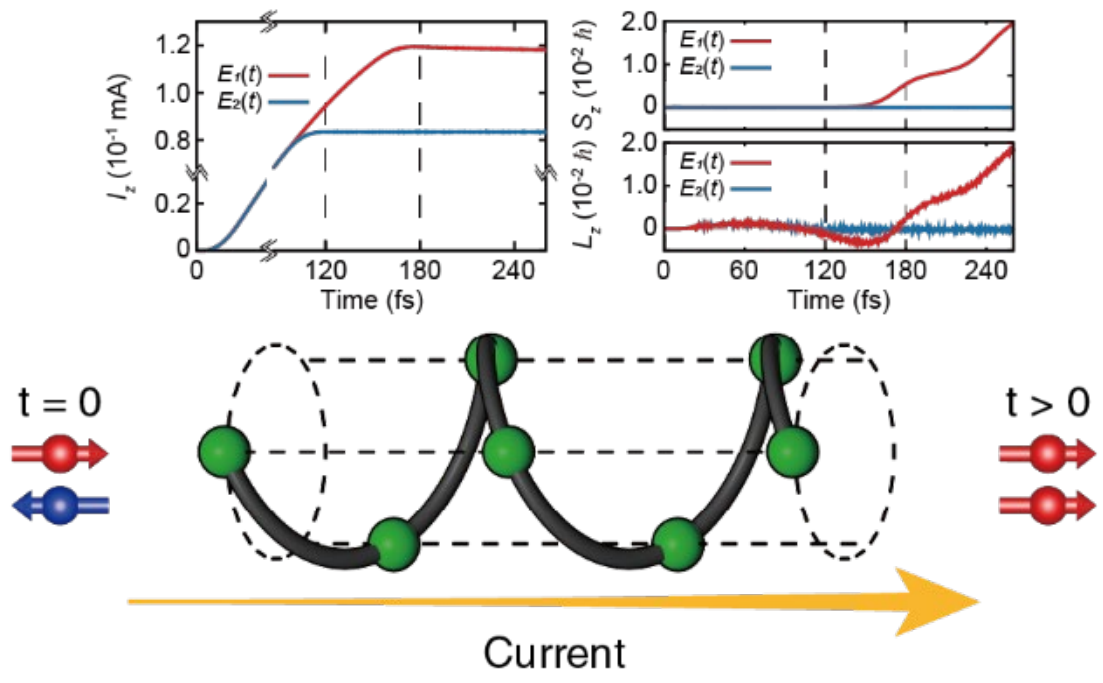

**Figure for ToC**

# Supplementary Information of Current-Driven Symmetry Breaking and Spin-Orbit Polarization in Chiral Wires

*Uiseok Jeong[1], Daniel Hill[2], Esmaeil Taghizadeh Sisakht[1], Binghai Yan[3], Angel Rubio[4], Carsten A. Ullrich[2], Noejung Park[1,4]**

[1]Department of Physics, Ulsan National Institute of Science and Technology(UNIST), UNIST-gil 50, Ulju-gun, 44919, Ulsan, Republic of Korea.

[2]Department of Physics and Astronomy, University of Missouri, 701 S College Ave, Columbia, 65211, Missouri, USA.

[3]Department of Physics, Pennsylvania State University, 201 Old Main, University Park, 16802, Pennsylvania, USA.

[4]Max Planck Institute for the Structure and Dynamics of Matter, Luruper Ch 149, Hamburg, 22607, Hamburg, Germany.

**Supplementary Note 1**: Screw-rotation symmetry operator and the symmetry breaking by the external field along the wire axis

Here, we explicitly demonstrate how the presence of an external electric field along the screw axis direction breaks the screw rotation symmetry. For the screw operator defined in Eq. 1 of the main text, the rotation and translation operators are defined as

$$\begin{aligned}\hat{R}_{2\pi/3} &= \exp\left(-\frac{i}{\hbar}\hat{L}_z\frac{2\pi}{3}\right), \quad \hat{L}_z = -i\hbar\frac{\partial}{\partial\varphi} \\ \hat{T}_{a/3} &= \exp\left(-\frac{i}{\hbar}\hat{p}_z\frac{a}{3}\right), \quad \hat{p}_z = -i\hbar\frac{\partial}{\partial z}.\end{aligned} \quad \text{Eq. S1}$$

Suppose we have a given screw-symmetric system:

$$\begin{aligned}\hat{H} &= \frac{1}{2m}\hat{\mathbf{p}}^2 + V(\mathbf{r}) \\ \hat{Q}^{-1}\hat{H}\hat{Q} &= \hat{H}.\end{aligned} \quad \text{Eq. S2}$$

Now, we introduce a static external electric field $\mathbf{E} = E\hat{z}$ along the screw axis. In the length gauge, the Hamiltonian becomes

$$\hat{H}_L = \frac{1}{2m}\hat{\mathbf{p}}^2 + V(\mathbf{r}) + eEz. \quad \text{Eq. S3}$$

As the rotation $\hat{R}$ never affect the *z*-coordinate and the translation $\hat{T}$ shifts *z* by $a/3$, we have:

$$\hat{Q}^{-1}z\hat{Q} = z + \frac{a}{3}. \quad \text{Eq. S4}$$

Therefore,

$$\hat{Q}^{-1}\hat{H}_L\hat{Q} = \frac{1}{2m}\hat{\mathbf{p}}^2 + V(\mathbf{r}) + eE\left(z + \frac{a}{3}\right) = \hat{H}_L + \frac{eEa}{3}. \quad \text{Eq. S5}$$

Consequently, the screw symmetry is broken by the axially biased electric field.

$$\hat{Q}^{-1}\hat{H}_L\hat{Q} \neq \hat{H}_L . \quad \text{Eq. S6}$$

**Supplementary Note 2**: Conservation of the total energy after the field is turned off

Suppose that the external electric field was applied only within a limited time duration and turned off at $t = t_2$. Here, we show that the total energy of the system is conserved for $t > t_2$. Consider the many-body Hamiltonian as follows,

$$\begin{aligned}
\hat{H} &= \frac{1}{2m}\int d^3\mathbf{r}\hat{\psi}^*(\mathbf{r})\left(-i\hbar\nabla + \frac{e}{c}\mathbf{A}\right)^2 \hat{\psi}(\mathbf{r}) + \int d^3\mathbf{r}\hat{\psi}^*(\mathbf{r})U(\mathbf{r})\hat{\psi}(\mathbf{r}) \\
&\quad + \frac{1}{2}\sum_{\alpha\beta}\iint d^3\mathbf{r}d^3\mathbf{r}'\hat{\psi}^{?}_{\alpha}(\mathbf{r})\hat{\psi}^{\dagger}_{\beta}(\mathbf{r}')W(\mathbf{r},\mathbf{r}')\hat{\psi}_{\beta}(\mathbf{r}')\hat{\psi}_{\alpha}(\mathbf{r}) \\
&= \int d^3\mathbf{r}\hat{\psi}^+(\mathbf{r})\frac{(-i\hbar\nabla)^2}{2m}\hat{\psi}(\mathbf{r}) + \frac{e}{c}\int d^3\mathbf{r}(-i\hbar\nabla)\cdot\mathbf{A} + \frac{e^2}{2mc^2}\int d^3\mathbf{r}\hat{\rho}(\mathbf{r})\mathbf{A}^2 + \int d^3\mathbf{r}\hat{\rho}(\mathbf{r})U(\mathbf{r}) \\
&\quad + \frac{1}{2}\sum_{\alpha\beta}\iint d^3\mathbf{r}d^3\mathbf{r}'\hat{\psi}^{?}_{\alpha}(\mathbf{r})\hat{\psi}^{\dagger}_{\beta}(\mathbf{r}')W(\mathbf{r},\mathbf{r}')\hat{\psi}_{\beta}(\mathbf{r}')\hat{\psi}_{\alpha}(\mathbf{r})
\end{aligned} \quad \text{Eq. S7}$$

Consider the time evolution of the many-body state ket.

$$\left|\Psi(t)\right\rangle = T\exp\left(-\frac{i}{\hbar}\int_0^t \hat{H}(t')dt'\right)\left|\Psi(t=0)\right\rangle \quad \text{Eq. S8}$$

For $t > t_2$, $\mathbf{A}(t)$ becomes constant, and thus the Hamiltonian is constant and the time-ordering is not necessary.

$$\left|\Psi(t)\right\rangle = \exp\left(-\frac{i}{\hbar}(t-t_2)\hat{H}(t_2)\right)\left|\Psi(t_2)\right\rangle \quad \text{Eq. S9}$$

Note, the time-evolution unitary operator commutes with the Hamiltonian, and the energy expectation from the time-evolving state for $t > t_2$ can be calculated as follows:

$$E(t) = \left\langle \Psi(t) \middle| \hat{H} \middle| \Psi(t) \right\rangle = \left\langle \Psi(t_2) \middle| e^{i\hat{H}(t-t_2)} \hat{H} e^{-i\hat{H}(t-t_2)} \middle| \Psi(t_2) \right\rangle = E(t_2) \quad \text{Eq. 10}$$

**Supplementary Note 3**: Time-evolution with the density-functional Hamiltonian of the given ground-state charge density

The exchange-correlation(XC) potential in the DFT Hamiltonian is, in principle, devised to describe the effects of interacting many electrons. Here, in Fig. S1, to separate the one-body effect from the collective responses of the interacting electrons, we repeated the same time evolution starting from the given initial charge density ( $\rho_0$ ), but without self-consistently updating either the charge density or the corresponding exchange–correlation potential (*i.e.*, with the density-functional potential held fixed). The orbital response obtained with this given fixed density-functional Hamiltonian exhibits mostly similar behavior to that from the full self-consistent time evolution. However, the time profiles of the responses of spin are drastically different: it is remarkably noticeable that the spin never emerges in this one-body time-evolution with the given fixed Hamiltonian. As discussed in the main text, this implies that orbital dynamics represent immediate one-body responses. On the other hand, the spin dynamics are based on a many-body nature.

**Supplementary Note 4**: Effect of chirality and spin–orbit coupling on momentum conversion dynamics

To further investigate the underlying mechanism of linear-to-angular momentum conversion, we additionally consider a non-chiral linear chain of $SnH_2$, as depicted in the top panel of Fig. S2(a). In order to disentangle the respective roles of chirality and SOC in driving momentum conversion processes under pure current flow, here the momentum dynamics of the non-chiral wire are compared with those of the chiral Se trigonal with or without SOC. This comparative study reveals that, even without SOC, the chirality alone can induce linear-to-angular momentum conversion. The external electric field was applied axially with a finite duration time of 180 fs, as used in Fig. 5 in the main text (the same electric field as $E_2(t)$ in Fig. 5(a)). The resulting time profiles of total energies, charge currents, orbital angular momenta, and spin angular momenta are displayed in Fig. S2(b, c, d, e), respectively. In agreement with the cases of Fig. 5 in the main text, the total energy is well conserved after the field is turned off. Note that the linear $SnH_2$ does not exhibit any spin dynamics. However, irrespective of the presence of the SOC, the chiral chains exhibit a decay of linear momentum, leading to the emergence of angular momenta: the SOC is indispensably required for spin dynamics; when the SOC is removed, the orbital angular momentum solely takes the role in the momentum conversion from linear to angular.

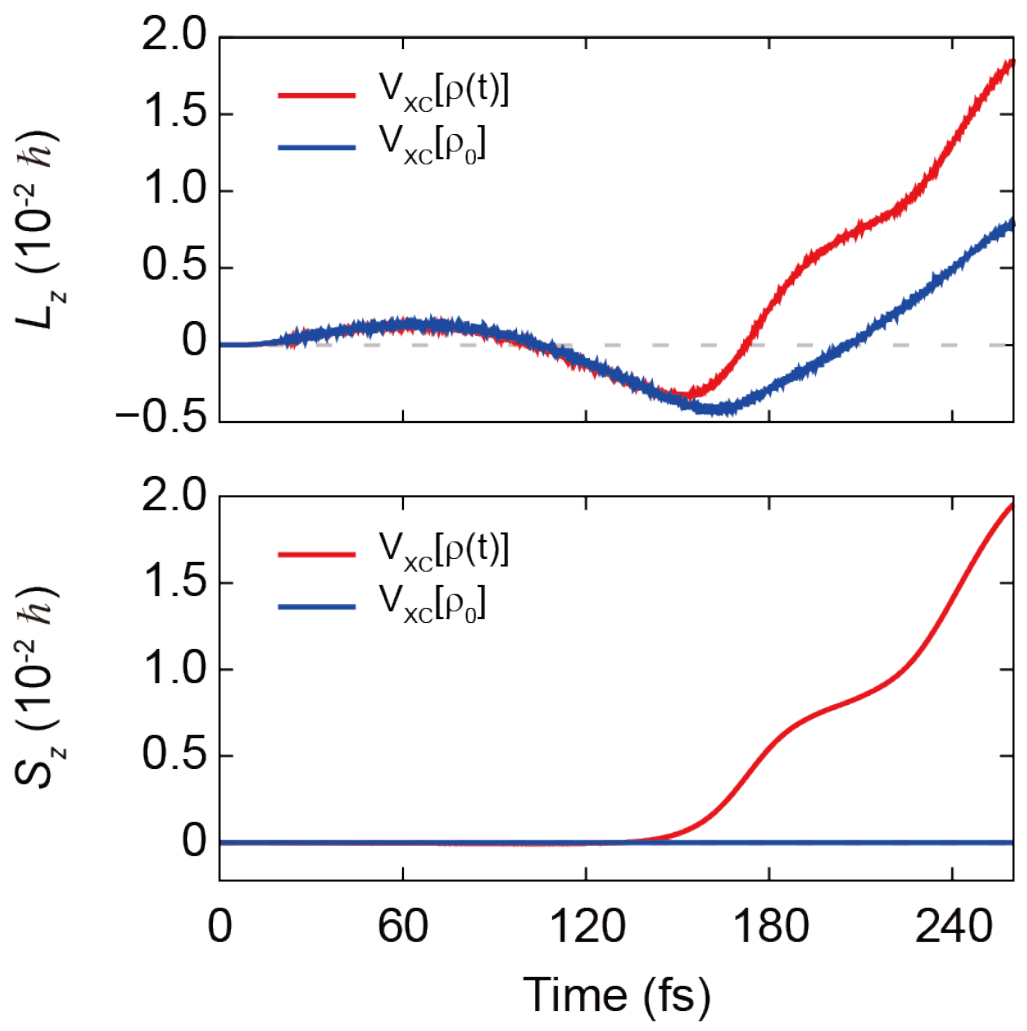

Figure S1. The effect of self-consistency in the exchange-correlation potential on orbital ( $L_z$ ) and spin ( $S_z$ ) dynamics. $\rho_0$ and $\rho(t)$ represent the frozen ground-state charge density and the self-consistently time-evolving density, respectively.

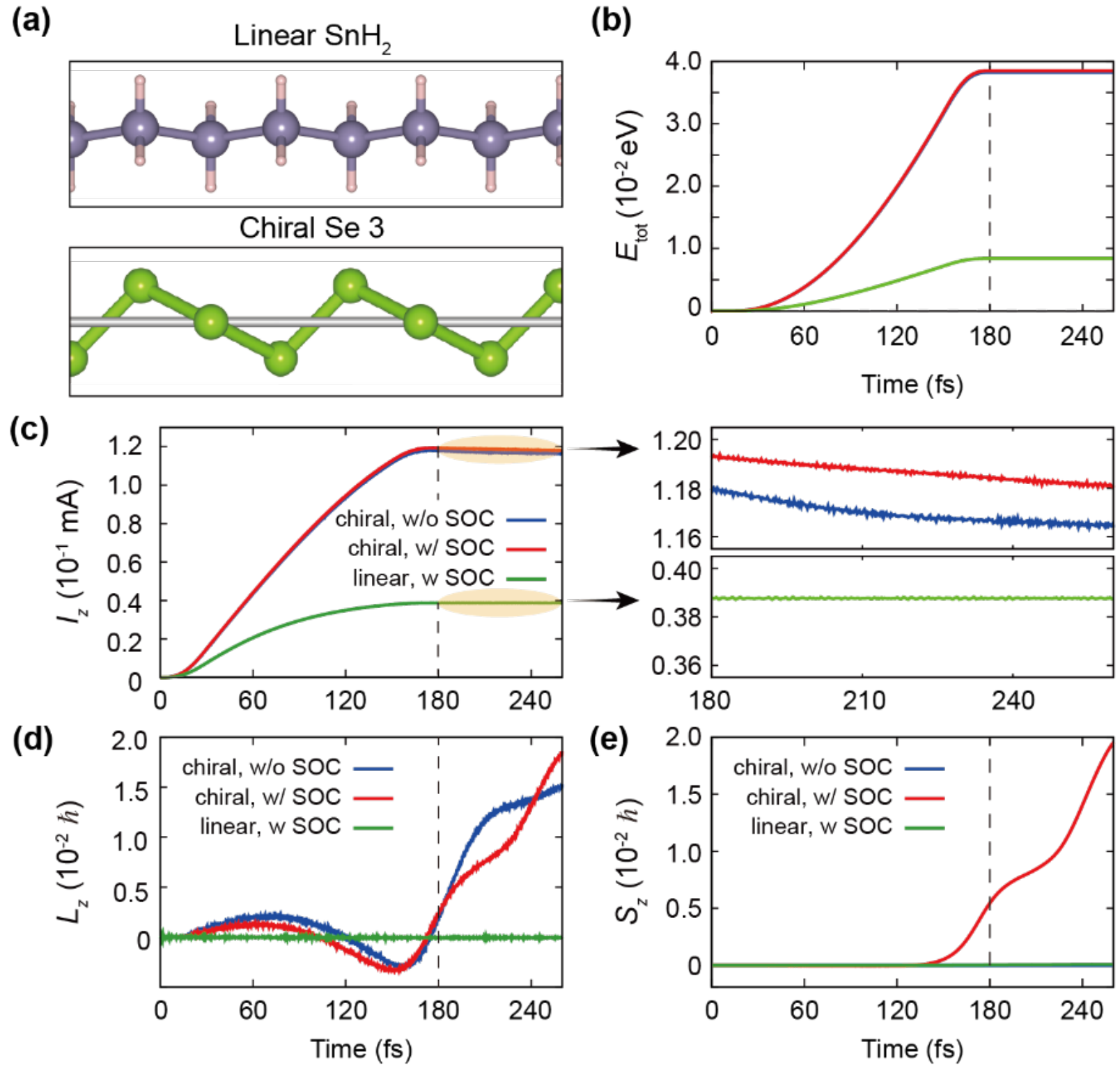

Figure S2. Comparison of linear-to-angular momentum conversion in chiral and non-chiral chains under pulsed electric fields. (a) Atomic configurations of non-chiral $SnH_2$ and chiral

trigonal Se wires. Time profiles of (b) total energies, (c) charge currents, (d) orbital angular momenta, and (e) spin angular momentum in response to the electric field persist up to 180 fs, as illustrated by the red line in Fig. 5(a) in the main text. To quantify the role of SOC, we additionally simulated the chiral Se wire with the SOC term removed from the Hamiltonian.

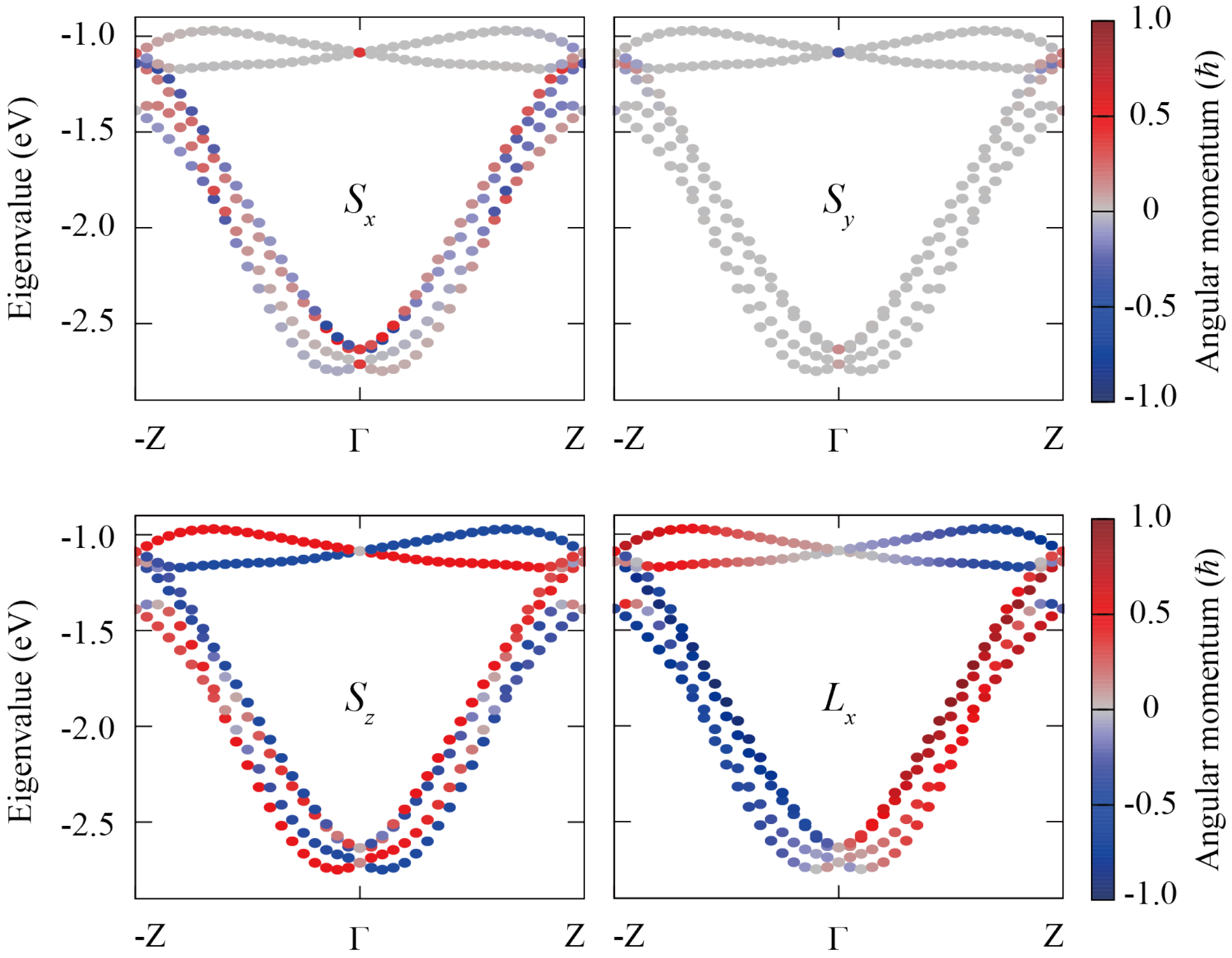

Figure S3. The electronic structure and spin-orbital texture of the Se trigonal one-dimensional wire with the screw symmetry being released by the y-directional perpendicular bias field.

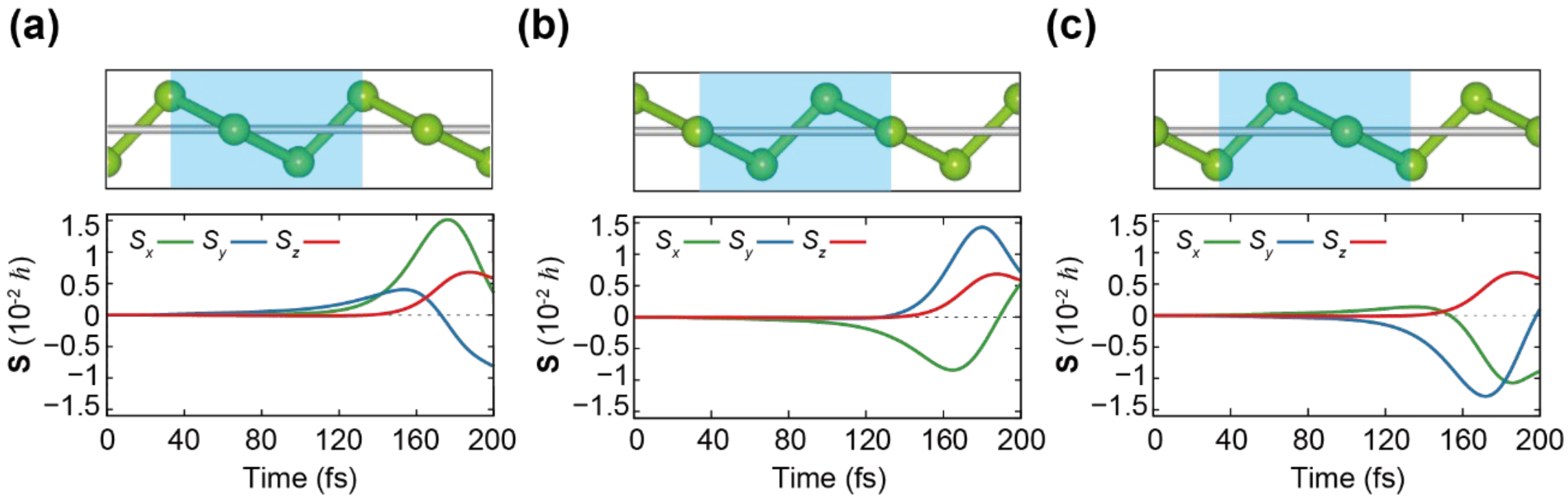

Figure S4. Arbitrariness of the spin angular momentum expectation value depending on the choice of unit cell: (a–c) Results for unit cells rotated by 0, $2\pi/3$, and $4\pi/3$, respectively.

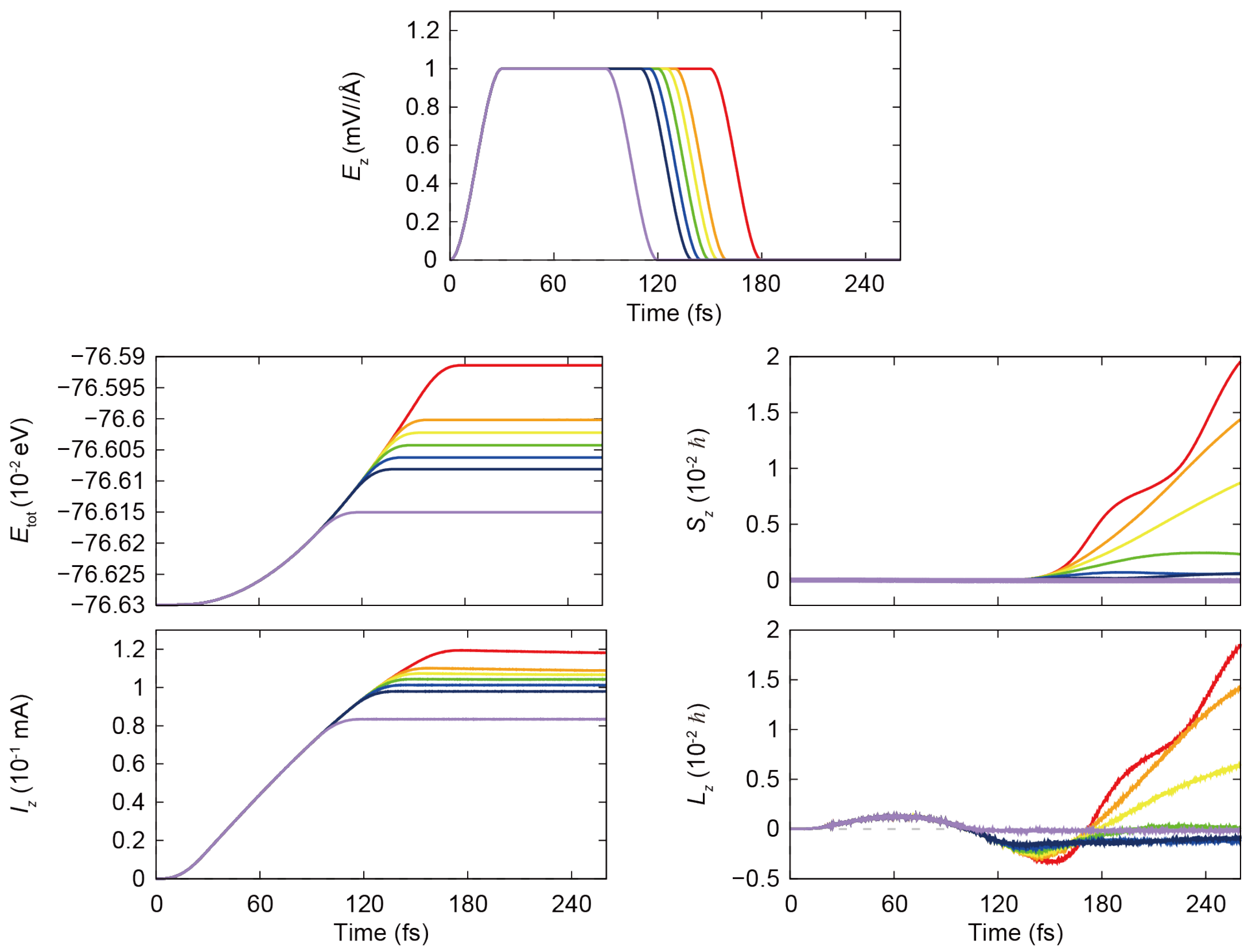

Figure S5. Time evolution of total energy, current, spin, and orbital angular momentum for various field durations, following the analysis presented in Fig. 5 of the main text.

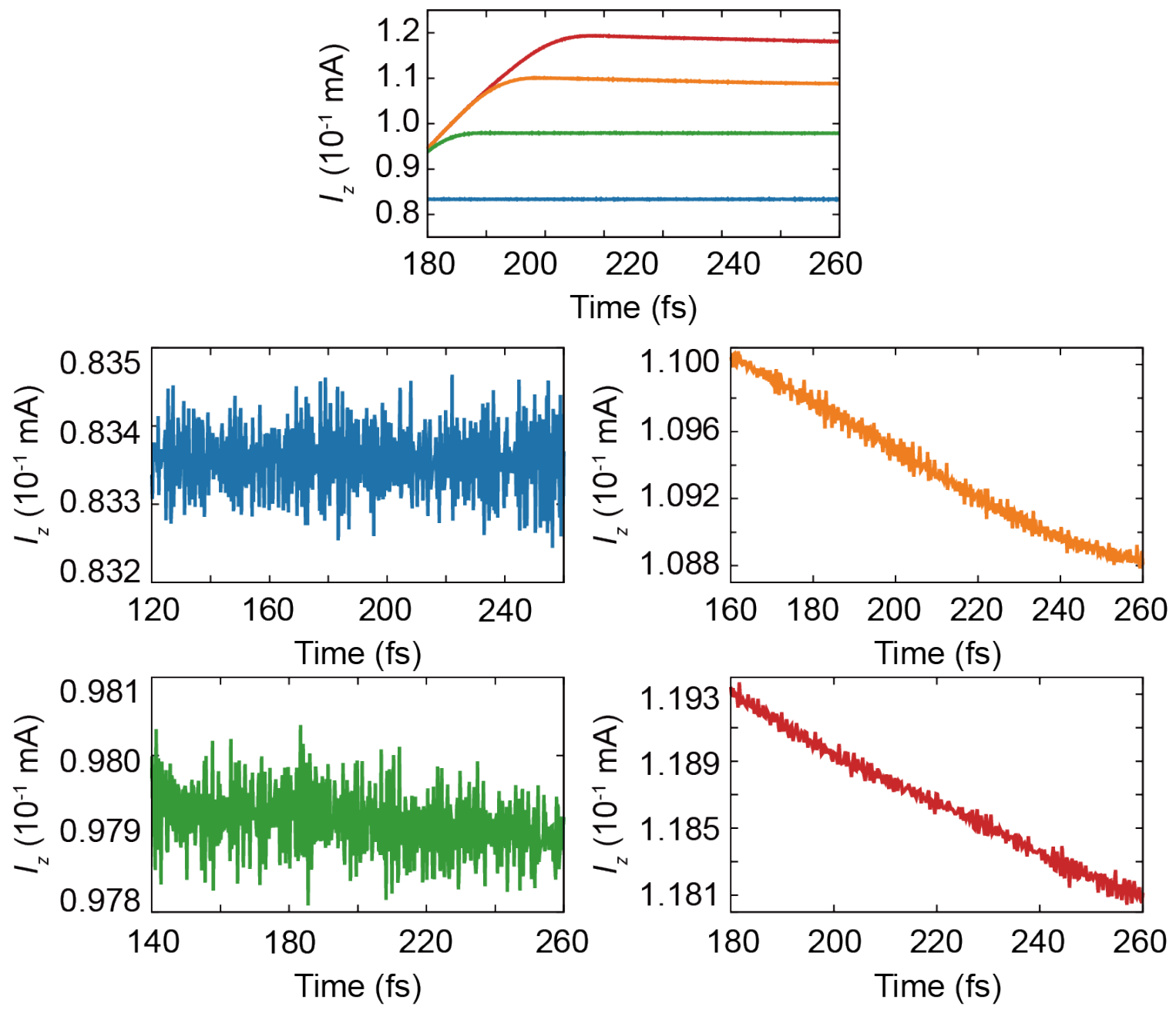

Figure S6. Magnified view of the current after the field duration. No current decay is observed for field durations of 120 fs and 140 fs, while noticeable decay occurs for durations of 160 fs and 180 fs.